\documentclass[10pt]{article}
\usepackage[]{geometry}
\usepackage[utf8]{inputenc}
\usepackage[super]{nth}
\usepackage[T1]{fontenc}

\usepackage{physics,amsmath,siunitx,amsfonts,amssymb,csquotes,hyperref,mathtools,lmodern,amsfonts,microtype,xcolor,booktabs,graphicx,moreverb,multicol,float,nameref,scrpage2,gensymb,tabu}

\ofoot{\pagemark}
\graphicspath{{ressources/}}

\author{Jonas Zeuner}
\title{Integrated Heralded Controlled-NOT Gate for Polarization-Encoded Qubits}

\usepackage{tikz}
\usepackage[backend=bibtex,style=nature,maxbibnames=5]{biblatex}
\bibliography{bibliography.bib} 

\usepackage{notoccite}

\begin{document}

\newcommand{\newtitle}[1]{
\begin{center}
{\textbf {\fontsize{14pt}{16.8pt}\selectfont #1}}
\end{center}
}

\newcommand{\authors}[1]{
\begin{center}
\textbf{\textit{\fontsize{12pt}{14.4pt}\selectfont #1}}
\end{center}
}

\newcommand{\newaffiliations}[1]{
\begin{center}
\textit{\fontsize{10pt}{12pt}\selectfont #1}
\end{center}
}

\setlength\oddsidemargin{-.2cm}
\setlength\evensidemargin{-1.1cm}
\setlength\textwidth{17.5cm}
\setlength\topmargin{-2cm}
\setlength\textheight{24cm}


\thispagestyle{empty}

\newtitle{Integrated-optics heralded controlled-NOT gate for polarization-encoded qubits} 

\authors{Jonas Zeuner$^1$*, Aditya N. Sharma$^{1}$, Max Tillmann$^{1}$, René Heilmann$^{2,3}$, Markus Gräfe$^{2,3}$, Amir Moqanaki$^{1}$, Alexander Szameit$^{2}$ and Philip Walther$^{1}$ }
\newaffiliations{
$^1$Faculty of Physics, University of Vienna, Boltzmanngasse 5, 1090 Vienna, Austria,\\
$^2$Institute for Physics, University of Rostock, Albert-Einstein-Stra\ss e 23, 18059 Rostock, Germany,\\
$^3$Institute of Applied Physics, Abbe Center of Photonics, Friedrich Schiller University Jena, Albert-Einstein-Stra\ss e 15, 07747 Jena, Germany
}

\noindent\textbf{
Recent progress in integrated-optics technology has made photonics a promising platform for quantum networks and quantum computation protocols. Integrated optical circuits are characterized by small device footprints and unrivalled intrinsic interferometric stability. Here, we take advantage of femtosecond-laser-written waveguides' ability to process polarization-encoded qubits and present an implementation of a heralded controlled-NOT gate on chip. We evaluate the gate performance in the computational basis and a superposition basis, showing that the gate can create polarization entanglement between two photons. Transmission through the integrated device is optimized using thermally-expanded-core fibers and adiabatically reduced mode-field diameters at the waveguide facets. This demonstration underlines the feasibility of integrated quantum gates for all-optical quantum networks and quantum repeaters.}

\vspace{1cm}



One of the most remarkable implications of quantum mechanics is the possibility of a machine that would dramatically outperform standard computers for certain tasks \cite{feynman1982}. Research groups around the world are pursuing a variety of approaches to develop such a quantum computer. Photonics has a rich history as a platform for fundamental quantum mechanics experiments \cite{hom,ViennaBell,NISTBell}, and it has developed into a competitive technology for quantum computing and quantum networks as well \cite{panPhotonicQC,waltherPhotonicQC,opticalQuantumRepeater}. One challenge facing the optical approach to quantum computing is that the traditional bulk-optics setups required to perform more complex experiments rapidly grow in size, and thus are challenging to stabilize. Integrated photonics offers a solution to this problem, promising intrinsic interferometric stability and the possibility of implementing a large number of quantum logic gates on a small monolithic chip. This technology has seen enormous progress in recent years \cite{femtoreview,englundreview,obriensiliconreview,gaasreview}, and may offer a realistic approach to realizing the complex circuits needed for scalable photonic quantum computing and quantum networks \cite{rudolphsOptimism}.
\par
An important feature of photonic qubits is their resistance to decoherence, even at room temperature. While photons’ limited interaction with the environment is an important advantage over matter systems in this context, it complicates the design of the two-qubit gates crucial for universal quantum computation (any quantum logic circuit can be realized using a combination of only single-qubit and two-qubit gates \cite{nielsenChuang}). However, the seminal work of Knill, Laflamme, and Milburn (KLM) showed that scalable linear optical quantum computing can be realized using only linear optical interferometers, single-photon sources, and single-photon detectors \cite{klm}. The two key concepts of the KLM scheme are (i) that the process of photon detection induces effective nonlinearites for two-photon gate operations, and (ii) that such gate operations can be achieved via the measurement of additional ``ancilla'' photons. Remarkably, it was shown that the success probability of linear-optical two-photon gates can be made arbitrarily close to unity by adding a sufficiently large number of ancilla photons. These findings inspired a large body of experimental and theoretical work \cite{onewayQC,onewayQCRudoplh}.
\par
The controlled-NOT (CNOT) gate is the quintessential two-qubit gate:~depending on the computational-basis state of the ``control'' qubit, the computational-basis state of the ``target'' qubit is either flipped or left unchanged (Fig.~\ref{fig:franson_cnot}). While a classical exclusive-OR (XOR) logic gate performs exactly this operation, a genuine CNOT gate must \textit{also} process control-target inputs that are quantum superpositions of the computational-basis states, maintaining coherence between them. In the latter case, the CNOT gate can be used to prepare a maximally entangled two-qubit state, or to unambiguously distinguish between all four possible Bell-state inputs: hence, the CNOT gate's importance to quantum computation evidently stems from its effect on superposition-state inputs.
\par
The earliest demonstrations of optical CNOT gates induced the KLM nonlinearity using destructive measurement of the photons carrying the control and target output qubits \cite{firstopticalcnot,kiesel2005linear,destructivebulkjeiji,whitedesructiveCNOTbulk}. This type of gate is \textit{unheralded}, since its successful operation can only be verified by measuring the output qubits themselves. Such gates are unsuitable for multi-step quantum logic circuits, where the output of one gate is used as input for the next: subsequent experiments used detection of additional ancilla photons to realize \textit{heralded} gates \cite{gasparoni2004firstcnot,panfirstnondestructiveCNOT,nondestructiveCNOTbulk,okamoto2011realization}. The technological progress of integrated quantum photonics has enabled demonstrations of unheralded integrated CNOT gates \cite{obriencnot,universallinearopticsobrien,crespicnot}, and recently also of a heralded CNOT gate \cite{universallinearopticsobrien}. In the latter experiment, the circuit was too small to fully implement a heralded CNOT gate and only the classical XOR logic operations were demonstrated.
\par
In this work, we demonstrate an integrated heralded CNOT gate using femtosecond-laser-written waveguides. In contrast to other integrated-optics platforms, which typically process path-encoded qubits (where two distinct waveguides represent logical states ``0'' and ``1'' of a single qubit), femtosecond-laser-written waveguides can additionally process polarization-encoded qubits. This feature allows us to halve the number of spatial modes used to implement the CNOT gate, and to adopt a flexible and modular scheme using high-quality free-space photon sources and detectors. A modular design could be of particular advantage  for all-optical quantum networks \cite{opticalQuantumRepeater} and delegated quantum computations \cite{blindquantumcomp}. In the long term, polarization processing capabilities offer other important benefits: simultaneous encoding of logic or ancilla qubits in multiple photonic degrees of freedom, such as path and polarization, can be used to increase the system's information-processing capacity or enable protocols that are otherwise impossible \cite{hbsa,pan10qubithyper}.
\par
A challenge facing femtosecond-laser-written waveguides and most other waveguide technologies is low coupling efficiency to and from the chip. The mode-field-diameter mismatch between conventional optical fibers and waveguide modes typically leads to high losses. Here, we significantly reduce these losses by coupling to the waveguides via thermally-expanded-core (TEC) fibers \cite{firsttec}, standard single-mode optical fibers (SMFs) whose mode-field diameters have been increased by an adiabatic thermal expansion process to match the waveguide mode-field diameter. In addition, the waveguide mode-field diameter is adiabatically reduced near the chip facet to better match the TEC fiber mode-field diameter \cite{jenaPaper}.
\par

\section*{Results}
\begin{figure}[h]
\center
\includegraphics[width=0.9\textwidth]{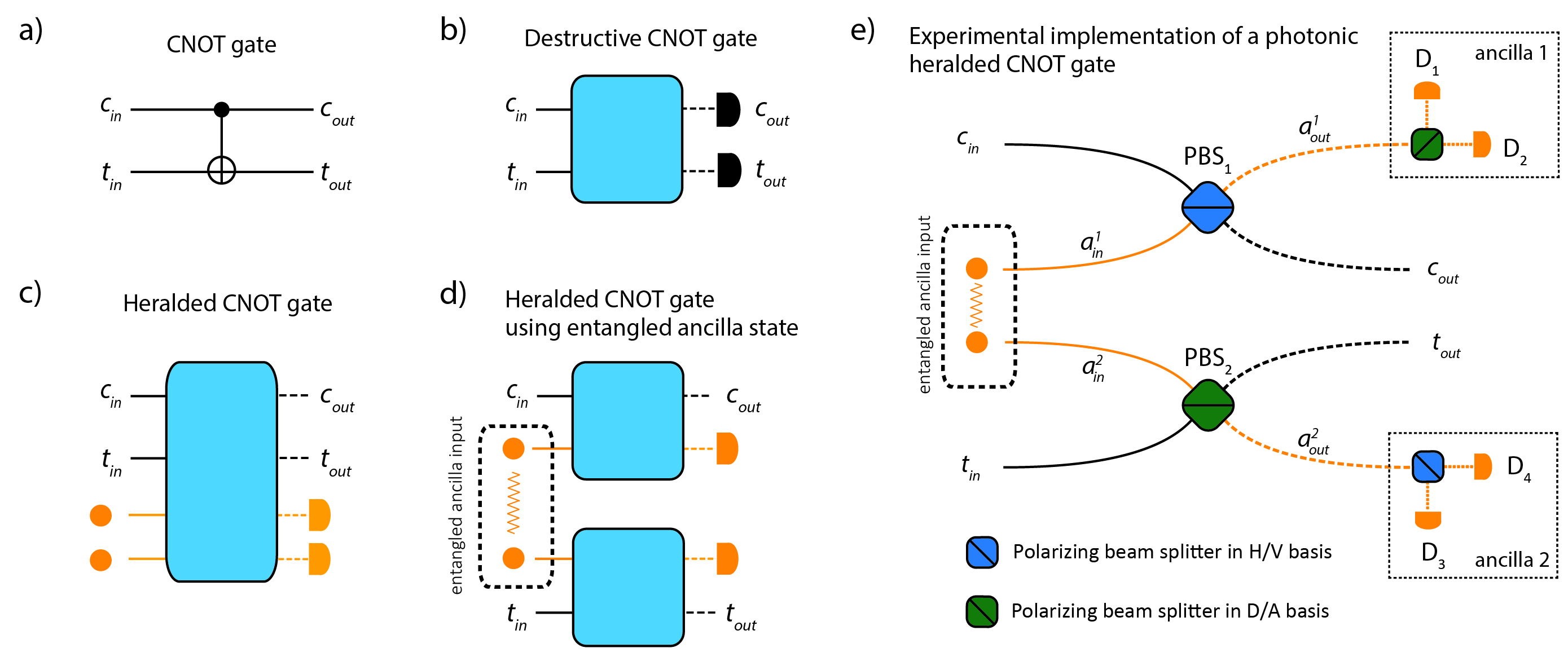}
\caption{\textbf{Schematic CNOT gates.} Black (orange) lines indicate logic (ancilla) qubits. Since the photonic gates in (b)-(e) rely on two-photon interference, in general there is no well-defined 1:1 correspondence between input and output photons: solid (dashed) lines indicate input (output) qubits. a) Logic circuit description of a CNOT gate. The gate processes ``control'' and ``target'' bits. The control input $c_\text{in}$ is transmitted directly to the control output $c_\text{out}$; the sum modulo 2 of $c_\text{in}$ and target input $t_\text{in}$ is transmitted to target output $t_\text{out}$. While this operation can be realized classically, a genuine CNOT gate must also preserve coherence of arbitrary control and target \textit{qubit} states. b) Unheralded photonic CNOT gate. The gate functions probabilistically and its success in any particular instance is ascertained by destructive measurement of the output control and target qubits. c) Heralded photonic CNOT gate. Two additional ancilla photons mediate the CNOT operation. Success of the probabilistic gate operation is heralded by detection of the output ancillas, so the control and target outputs are available for subsequent logic operations. d) Heralded photonic CNOT gate using a maximally entangled ancilla state. The special case in which the ancilla photons in (c) are initially entangled allows for improved success probability. e) An implementation of the scheme shown in (d) using polarization qubits. With probability 1/4, exactly one photon is detected in each of the modes $a^1_\text{out}$ and $a^2_\text{out}$, heralding successful gate operation \cite{pittman2001probabilisticcnot}. In this case, the state $\ket{c_\text{out},t_\text{out}}$ matches the desired CNOT output up to a known Pauli rotation: depending on the ancilla two-photon polarization measurement outcome, one of four feed-forward unitaries must be applied to the control and target photons after a successful gate operation (not shown).}
\label{fig:franson_cnot}
\end{figure}

\textbf{Working principle of the heralded CNOT gate.} 
Our CNOT gate operates on photon polarization. We define horizontal ($H$) and vertical ($V$) polarization to correspond to the logic states ``0'' and ``1''; we will also refer to the single-photon polarization states $\ket{D}=(\ket{H}+\ket{V})/\sqrt{2}$, $\ket{A}=(\ket{H}-\ket{V})/\sqrt{2}$, $\ket{L}=(\ket{H}+i\ket{V})/\sqrt{2}$, and $\ket{R}=(\ket{H}-i\ket{V})/\sqrt{2}$. To realize a heralded gate, we use a maximally entangled pair of ancilla photons in the Bell state $\ket{\Psi^-}=(\ket{H,V}-\ket{V,H})/\sqrt{2}$ (Fig.~\ref{fig:franson_cnot}d); we will also refer to the other two-photon polarization Bell states $\ket{\Phi^\pm}=(\ket{H,H}\pm\ket{V,V})/\sqrt{2}$ and $\ket{\Psi^+}=(\ket{H,V}+\ket{V,H})/\sqrt{2}$.
\par
Following the proposal of Ref. \cite{pittman2001probabilisticcnot}, our experiment uses two polarizing beam splitters (PBSs) in mutually unbiased bases (Fig.~\ref{fig:franson_cnot}e). The ideal gate operation is as follows: exactly four indistinguishable photons, are input to the gate in polarization state $\ket{c_{\text{in}},t_{\text{in}} }\otimes\ket{a^1_\text{in},a^2_\text{in}}$, where $\ket{c_{\text{in}},t_{\text{in}}}$ is the input state of the control and target qubits, and $\ket{a^1_\text{in},a^2_\text{in}}=\ket{\Psi^-}$ is the state of the input ancilla photons (while Ref. \cite{pittman2001probabilisticcnot} instead uses $\ket{\Phi^+}$ as the ancilla state, this difference has no effect on the essential functionality of the gate, and simply changes the output by a known Pauli rotation). A CNOT operation is performed on $\ket{c_{\text{in}},t_{\text{in}}}$ conditional on simultaneous detection of two output ancilla photons, exactly one at detectors $D_1,D_2$ and exactly one at detectors $D_3,D_4$, which happens with probability 1/4. Such a two-fold coincidence detection heralds successful gate operation and the presence of the two remaining photons in the output control-target state $\ket{c_{\text{out}},t_{\text{out}}}$. However, to obtain the desired CNOT output state, one of four Pauli rotations must be performed on this output control-target state, depending on the polarization measurement outcome for the output ancilla state $\ket{a^1_\text{out},a^2_\text{out}}$. The output control and target qubits need not be measured and can be used as input for subsequent logic operations.
The two PBSs are physically identical; however, polarization rotations are applied to the photons before entering and after exiting the second PBS such that it acts in a rotated basis.

\vspace*{1cm}

\begin{figure}[h]
\center
\includegraphics[width=0.75\textwidth]{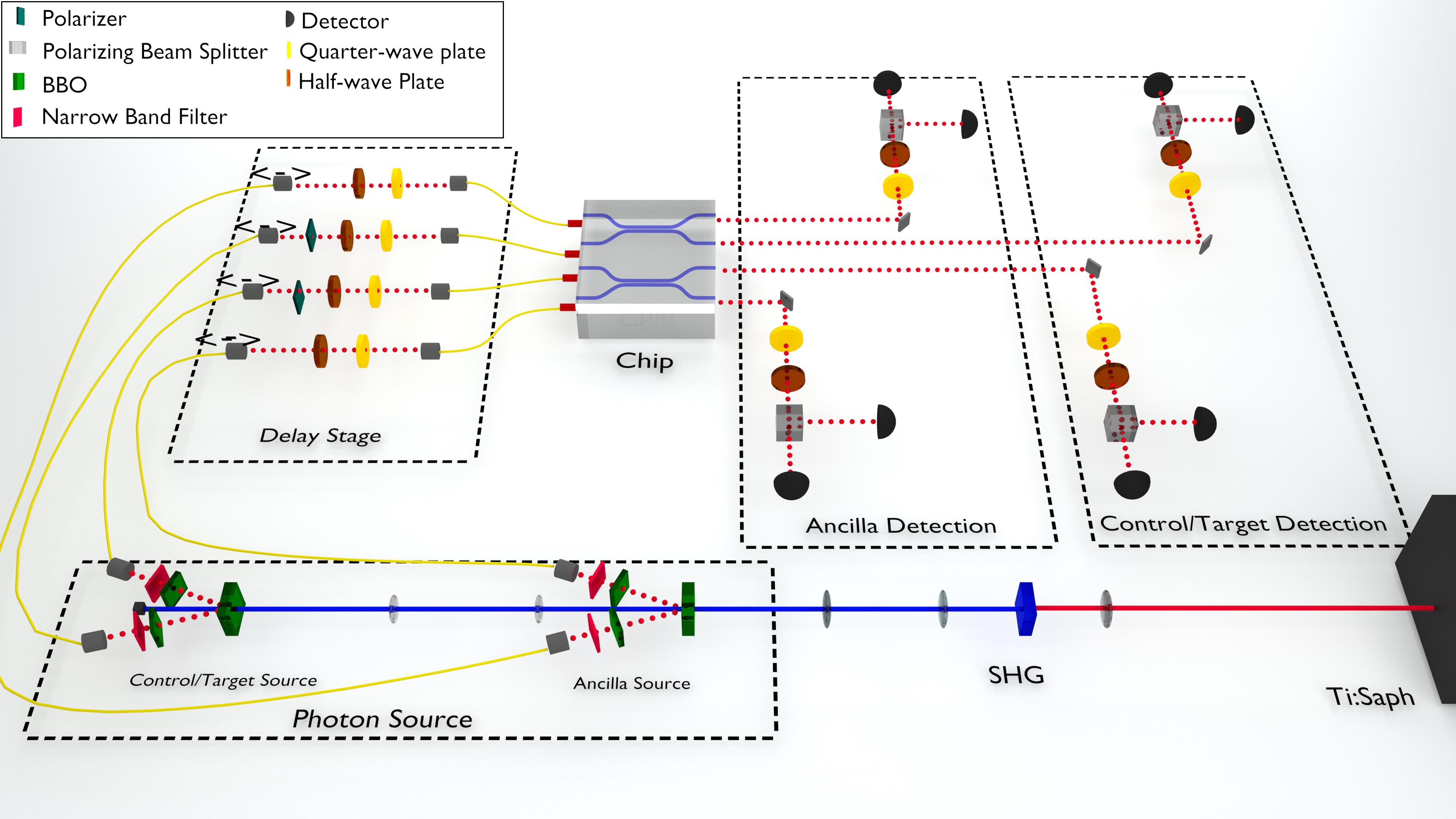}
\caption{\textbf{Experimental setup.} Frequency-doubled pulses from a Ti:sapph laser create photon pairs in two nonlinear $\beta$-barium borate crystals via degenerate type-II spontaneous parametric down conversion. Each crystal produces a maximally entangled polarization state. The photons from the first crystal are used as maximally entangled ancillas; the photons from the second crystal are also initially maximally entangled, and can be converted to any two-qubit control-target state using suitable projections and rotations. Before entering the waveguide, the photons are synchronized in the delay stage:  the free-space length of the delay stage is tuned independently for each photon using servo motors, as indicated by the double-headed arrows. The TEC fibers coupled to the chip have thermally expanded cores near their exit facets (indicated by the red tips), to increase mode overlap with the waveguides. After interfering on the chip, the four output photons are collimated with a single lens at the output facet of the chip (omitted in the figure), and sent to the detection stage for polarization analysis. Each photon passes through a quarter-waveplate, a half-waveplate, and a polarizing beam splitter before being coupled to multi-mode fiber (omitted in the figure) and sent to an avalanche photodiode.  Feed-forward is simulated in post-processing: the control-target measurement outcome is reassigned to one of four values depending on the ancilla measurement outcome.}
\label{fig:setup}
\end{figure}

\noindent\textbf{Experimental Setup} As shown in Fig.~\ref{fig:setup}, we generate the control, target, and ancilla photons using degenerate, non-collinear type-II spontaneous parametric down-conversion (SPDC). A Ti:sapphire laser (Coherent Chameleon Ultra II) is used to generate pulses ($\SI{150}{fs}$ duration, $\SI{80}{MHz}$ repetition rate, $\SI{3.8}{W}$ average power) at a wavelength of $\SI{789}{nm}$, which then undergo second harmonic generation in a $\SI{5}{mm}$-thick lithium triborate (LBO) crystal. The resulting beam is spatially filtered and used to pump two separate $\SI{2}{mm}$-thick $\beta$-barium borate (BBO) crystals, probabilistically producing two polarization-entangled photon pairs. Half-waveplates (HWPs) and $\SI{1}{mm}$-thick BBO crystals of the same cut angle as the SPDC crystals are placed in all four photons' paths to compensate for temporal and spatial walk-off, such that both sources produce pairs in polarization state $\ket{\Psi^-}$ \cite{kwiat1995}. The first (second) SPDC source supplies the ancilla (control and target) photons. Since we start with a maximally entangled control-target state, we can prepare any desired input $\ket{c_\text{in},t_\text{in}}$ using suitable single-photon projections and rotations. The pump power is attenuated to approximately $\SI{600}{mW}$ ($\SI{550}{mW}$) at the first (second) source's BBO crystal. All four photons pass through $\SI{3}{nm}$-bandwidth spectral filters centered at $\SI{789}{nm}$, and through single-mode optical fibers (SMFs) (Nufern 780-HP). This filtering results in polarization-entangled photons with high spectral and spatial purity (see Methods).

Adjustable free-space delays lines are used to synchronize the photons such that they all arrive at the chip simultaneously to within their coherence time of approximately \SI{300}{fs}. We couple from the SMFs to the chip via a \SI{127}{\micro\meter}-pitch v-groove array. The chip consists of two layers. The layer used for the CNOT gate couples the four input and output waveguides to two PBS structures. The second layer is used for alignment and contains two non-polarizing beam splitters (BSs), and several uncurved \enquote{calibration guides}. We use this second layer, which is positioned \SI{127}{\micro\meter} below the CNOT layer, to search for Hong-Ou-Mandel interference while tuning the relative time delays between the photons \cite{hom}. The fibers coupled to the chip are TEC SMFs, whose mode-field diameters have been increased from \SI{5}{\micro\meter} to \SI{10}{\micro\meter} in the region near their exit facets \cite{tec1,tec2,tec3}, boosting the expected overlap with the adiabatically compressed \SI{8x11}{\micro\meter} elliptical waveguide modes from 68\% to over 96\%, corresponding to a 41\% improvement over unmodified SMFs (see \textbf{Waveguide Details}). Although the maximal coupling efficiency we achieve in our experiment is only $(76\pm5)\%$, we indeed observe an increase of  $(46\pm10)\%$ per mode in coupling efficiency, consistent with the theoretical prediction. Moreover, we later discovered that improved polishing of the chip increases the coupling efficiency to $(90\pm 8)$\% , suggesting that our performance was limited either by surface roughness or by waveguides that terminated before reaching the chip facet.  All coupling efficiency errors cited here assume that for each fiber type we were able to achieve coupling within $5\%$ of the optimum, and that all input and output powers were measured to within $5\%$ of the actual values.  On-chip loss is on the order of \SI{0.3}{dB/cm} when there is no curvature in the waveguide mode. The overall transmission (from fiber in-coupling up to the avalanche photodiodes (APDs)) of the straight calibration guides was measured to be ($50\pm5$)\%; in each of the four modes used for the CNOT gate, we measured a transmisson of $(40\pm 5)$\%, most likely due to higher bending losses in the curved waveguide sections.
\par
After exiting the waveguides, the photons are sent to a detection stage. Each output mode of the waveguide is sent to a free-space motorized quarter-waveplate (QWP) and HWP followed by a PBS, enabling measurements in any polarization basis. Both outputs of each of these PBSs are coupled to APDs through multi-mode fibers. Four-photon polarization data is collected in the event of coincidence detection of photons in each of the modes $c_\text{out}, t_\text{out}, a^1_\text{out}, a^2_\text{out}$.
\par
\vspace*{1cm}
\noindent\textbf{Waveguide Details.} The chip used in this experiment was fabricated by direct femtosecond-laser writing \cite{firstfemtopaper,firstfemtoquantumpaper}: in this process, an ultrashort laser pulse (\SI{150}{fs}, \SI{0.5}{mJ}, \SI{100} {kHz}, \SI{800}{nm}) is tightly focused into a fused silica sample, causing nonlinear absorption and permanently increasing the refractive index in the focal volume  \cite{szameit2010discrete,szameitreview}. The guiding regions are traced out by a series of such pulses, and at $\SI{789}{nm}$ they support single modes with elliptical profiles (\SI{15x20}{\micro \meter}). This ellipticity is due to the intensity distribution of the writing laser beam waist and contributes to slightly different refractive indices for $H$ and $V$ ($\Delta n\approx10^{-5}$) \cite{szameitpbs}. At the facets of the chip, the minor and major diameters of the approximately elliptical mode are compressed to \SI{8x11}{\micro\meter} using an adiabatic writing technique, in order to increase overlap with the fiber mode \cite{jenaPaper}.
\par
\begin{figure}[h]
\center
\includegraphics[width=0.9\textwidth]{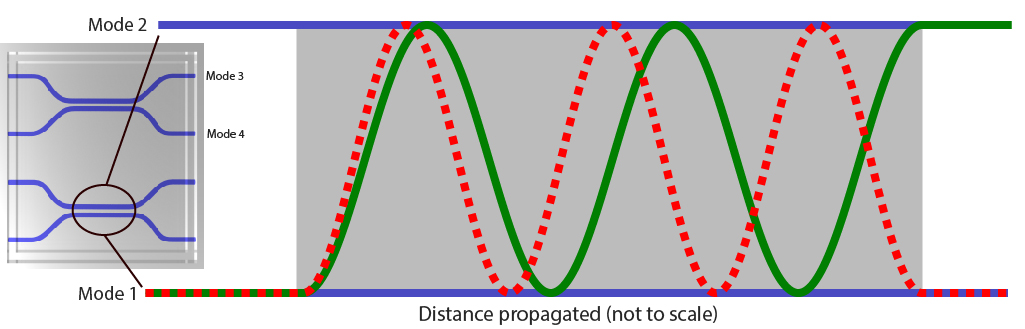}
\caption{\textbf{Schematic illustration of the polarization-dependent directional coupler.} Light initially propagating in one waveguide can evanescently couple to another at a rate determined by the spatial separation and refractive indices of the two waveguides. Thus, the probability amplitude for finding a photon initially in Mode 1 can oscillate between Modes 1 and 2. Due to the slight birefringence, the oscillation rates are slightly different for horizontal ($H$) and vertical ($V$) polarizations. The inset shows a schematic of the chip, which implements directional couplers for Modes 1 and 2, and Modes 3 and 4. The evanescent coupling regions for the directional couplers are designed such that all $V$-polarized light (dashed red line) remains in Mode 1 at the end of the region, while all $H$-polarized light (solid green line) is transferred to Mode 2, similar to a free-space PBS. A different choice of parameters can lead to equal probabilities for the two modes simultaneously for $H$ and $V$, producing a 50:50 BS like the ones we implement on the lower layer of the chip (not shown).}
\label{fig:chip_render}
\end{figure}

The two PBSs used for our CNOT gate are implemented using a scheme that exploits the waveguides' birefringence. Evanescent coupling between two guided modes can occur for separation distances ranging from $20-\SI{30}{\micro \meter}$, and different types of directional couplers can be constructed depending on the length of the coupling region and the relative position of the modes with respect to their ellipse axes \cite{szameitbs,osellamebs,osellamePBS}. Since the strength of the evanescent coupling is different for $H$ and $V$, it is possible to tune the length of the coupling region in such a way that $H$ returns entirely to the input mode while $V$ transfers entirely to the other mode. Such a structure exhibits the PBS behavior we require (Fig. \ref{fig:chip_render}). The same principle, with different parameters, is used to fabricate the BSs on the second layer of the chip. Due to birefringent dispersion, the optimal performance of all of these structures is wavelength-dependent: we measured an extinction of 50:1 for the PBS structures using our \SI{3}{\nano \meter}-bandwidth SPDC photons. Although this birefringence is central to the waveguide functionality, it also induces unknown phase shifts between $H$ and $V$, which simply correspond to local unitary rotations. This effect was compensated for in post-processing.

\begin{figure}[h]
\center
\includegraphics[width=0.9\textwidth]{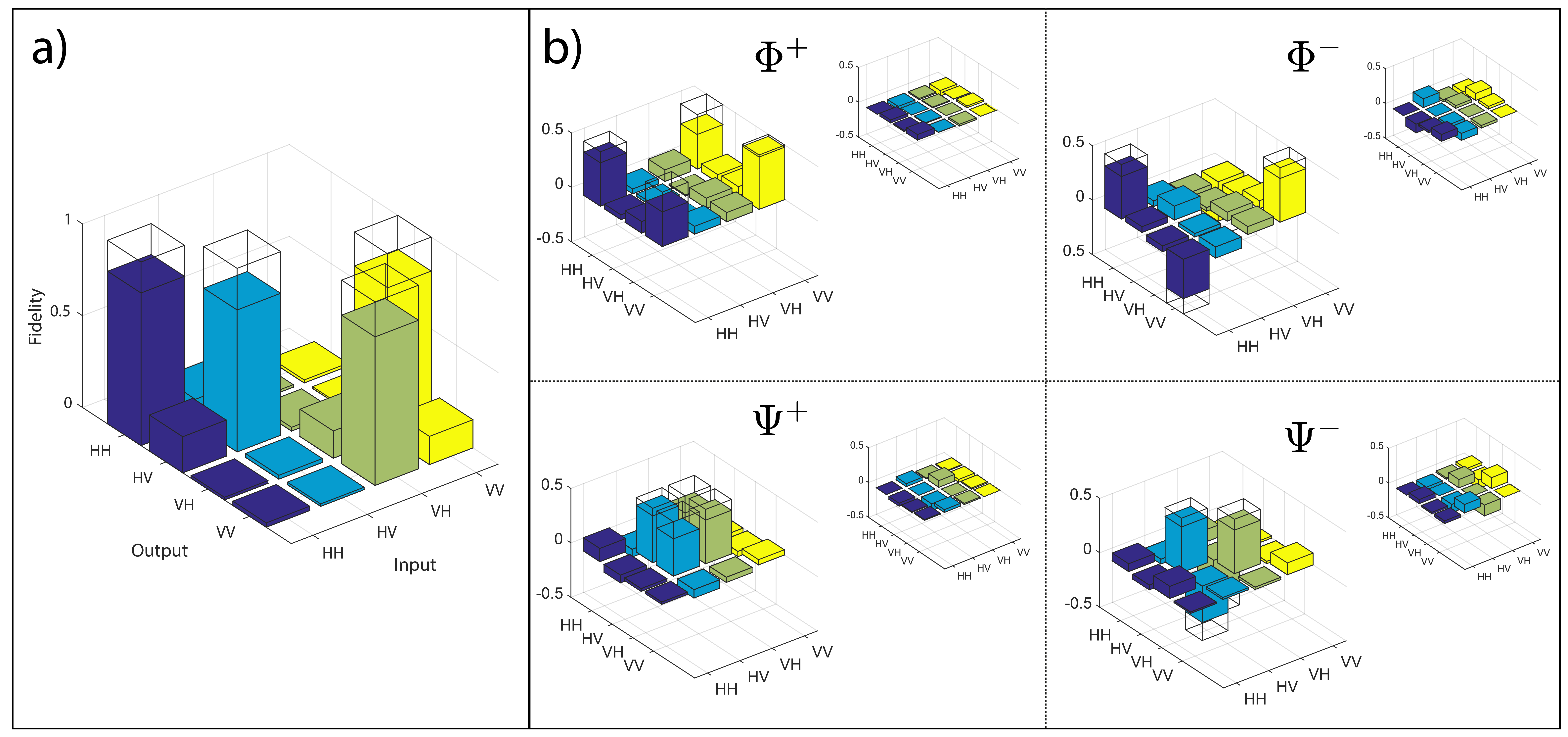}
\caption{\textbf{Truth table and heralded generation of Bell states.} a) Truth table showing the probability of each computational-basis output for each computational-basis input. The ideal gate behavior is indicated by the wireframe. The target qubit cross-talk is due to imperfect two-photon interference between partially distinguishable photons from our two sources. The overlap with the ideal truthtable is ($83.8\pm 2.6) \%$.  b) The real (large) and imaginary (small) parts of the density matrices for the four states produced by the CNOT, depending on which ancilla measurement outcome occurs.  The fidelity of these matrices with their respective target Bell states (indicated by the wireframe) are $76\%$ ($\ket
{\Phi^+}$), $76\%$ ($\ket{\Phi^-}$),$74  \%$ ($\ket{\Psi^+}$) and  $76\%$ ($\ket{\Phi^-}$) all with errors of $\pm 5\%$.  }
\label{fig:bellstates}
\end{figure}

\vspace*{1cm}
\noindent\textbf{Performance of the gate} It is natural to discuss the performance of a CNOT gate for two types of control-target inputs. In the first, the input state is one of the four computational basis states $\ket{HH}, \ket{HV}, \ket{VH}, \ket{VV}$, and the gate output can be replicated using only classical operations. We evaluated the computational-basis performance of our CNOT gate by measuring the truth table (Fig.~\ref{fig:bellstates}a), which requires four single-setting measurements: for each input state, we analyzed the output in the computational basis. Finally, in post-processing the data, we simulated feed-forward Pauli rotations, reassigning the measurement outcomes for control and target depending on which of the four possible measurement outcomes occurred for the two ancilla photons. The overlap of the measured truth table with the ideal one is $(83.8\pm 2.6)\%$.
\par
On the other hand, when the control input is a superposition of the computational-basis states, the gate's behavior necessarily demands a quantum-mechanical explanation. For the control-target input state $\ket{D,H}$, for example, the gate outputs a coherent superposition of the outputs for $\ket{H}$ and $\ket{V}$ control inputs giving the maximally entangled state $\ket{\Phi^+}$. Here, in contrast to the computational-basis case, a single measurement does not suffice to demonstrate the gate's behavior: for this input state, we collected data for nine measurement settings and reconstructed the two-qubit density matrix using quantum state tomography (QST) for each of the different ancilla measurement outcomes \cite{tomography}. 
To present  the density matrices in standard Bell-state form, we applied local unitaries in post-processing to compensate for the chip birefringence (Fig.~\ref{fig:bellstates}b). The four inferred states correspond to the four possible ancilla outcomes for a single input before feed-forward. These density matrices are presented quantitatively in the Supplementary Material.
\par
For both the computational-basis and superposition-basis cases, we subtracted from our data the rate of higher-order emissions leading to four-fold coincidence detection. It is important to note that this noise is a limitation of our SPDC source, not the gate. Moreover, with the rapid development of other types of single-photon sources, such as those based on quantum dots \cite{panbosonsampling,senellartQD}, it is likely this obstacle will soon be overcome. Noise subtraction is discussed further in \textbf{\nameref{sec:dataAnalysis}}. After subtraction, the four-fold coincidence rate was on the order of $\SI{100}{mHz}$.

\section*{Discussion}

 We have performed a complete demonstration of an integrated heralded CNOT, characterizing the gate operation for computational-basis and superposition-basis control qubit inputs. We used TEC fibers and adiabatic writing techniques to efficiently transfer high-fidelity polarization qubit states from free-space sources into an integrated-optic circuit. Taking advantage of femtosecond-laser-written waveguides' polarization control capabilities, we processed these qubits directly on chip.
\par
By adiabatically modifying the mode-field diameters of the SMFs and the waveguides, we increased the coupling efficiency between them, strongly improving on previous work \cite{maxwaveguide}. The measured increase in coupling efficiency matched mode-field overlap calculations; additionally, a recent measurement with improved polishing of the chip yielded an absolute coupling efficiency close to the theoretical prediction.
\par
Our work expands on the first demonstrations of unheralded integrated CNOT gates using  polarization- or path-encoding \cite{crespicnot,obriencnot,universallinearopticsobrien}, combining the desirable features of heralding, integration and polarization encoding. However, to make linear-optics quantum computation, all-optical quantum networks, and quantum repeaters a reality,  technological challenges remain to be overcome. Of particular importance are the development of near-deterministic single-photon sources and near-unit-efficiency single-photon detectors, as well as low-loss waveguides with active control of polarization and phase shifts.
\section*{Methods}
\label{sec:methods}

\subsection*{Detailed Gate Logic}
\label{sec:gatelogic}
We describe the gate in two parts, first addressing the interference at $\text{PBS}_1$ and then the interference at $\text{PBS}_2$, following the analysis in Ref. \cite{pittman2001probabilisticcnot}. Let us consider the input state
\begin{equation}
\ket{c_\text{in},t_\text{in}}=(\alpha \ket{H}+\beta \ket{V})\otimes(\gamma \ket{H}+\delta\ket{V}).
\end{equation}
The ancilla photons in modes $a^1_\text{in},a^2_\text{in}$ are always input in the maximally entangled polarization state
\begin{equation}
\ket{a^1_\text{in},a^2_\text{in}}=\ket{\Psi^-}=\frac{1}{\sqrt{2}} (\ket{H,V}-\ket{V,H}).
\end{equation}
$\text{PBS}_1$ encodes the input control photon's polarization state onto the joint state of the photons in $a^2_\text{in}$ and $c_\text{out}$ as follows (see Fig. \ref{fig:franson_cnot}): when exactly one photon arrives in each of the modes $a^1_\text{out}$ and $c_\text{out}$, the photons in modes $a^1_\text{out},a^2_\text{in},c_\text{out}$ are projected onto the state
\begin{align}
&\alpha \ket{HVH}_{a^1_\text{out},a^2_\text{in},c_\text{out}}+ \beta \ket{VHV}_{a^1_\text{out},a^2_\text{in},c_\text{out}}\nonumber\\
&=\frac{1}{\sqrt{2}}(\ket{D}+\ket{A})_{a^1_\text{out}}\otimes\alpha\ket{VH}_{a^2_\text{in},c_\text{out}}+\frac{1}{\sqrt{2}}(\ket{D}-\ket{A})_{a^1_\text{out}}\otimes\beta\ket{HV}_{a^2_\text{in},c_\text{out}}.
\end{align}
When the ancilla photon in $a^1_\text{out}$ collapses onto polarization state $\ket{D}$, the photons in $a^2_\text{in},c_\text{out}$ are projected onto the polarization state
\begin{equation}
\ket{\psi_{\text{PBS}_1}}=\alpha \ket{VH}_{a^2_\text{in},c_\text{out}} + \beta \ket{HV}_{a^2_\text{in},c_\text{out}},
\label{case1}
\end{equation}
and when the photon in $a^1_\text{out}$ is instead found in polarization state $\ket{A}$, the photons in $a^2_\text{in},c_\text{out}$ are projected onto
\begin{equation}
\mathbb{I}_2\otimes\sigma_z\ket{\psi_{\text{PBS}_1}}=\alpha \ket{VH}_{a^2_\text{in},c_\text{out}}
- \beta \ket{HV}_{a^2_\text{in},c_\text{out}}.
\label{case2}
\end{equation}
Thus, when exactly one photon arrives at detectors $D_1,D_2$, the input control qubit has successfully been encoded onto the state shared by $a^2_\text{in}$ and $c_\text{out}$, and the probability of such an event is $\frac{1}{2}$. Depending on which of the detectors $D_1$ or $D_2$ clicks, a feed-forward operation may need to be applied to the control photon as indicated in (\ref{case2}).
\par
Although the interference at $\text{PBS}_2$ works essentially the same way, it is helpful to take a different perspective and to consider two examples. If the photon in $a^2_\text{in}$ is in state $\ket{H}$, detection of exactly one photon at $D_3,D_4$, which occurs with probability $\frac{1}{2}$, projects the gate output onto
\begin{equation}
\frac{1}{\sqrt{2}}(\ket{V}_{a^2_\text{out}}\otimes(\gamma\ket{H}+\delta\ket{V})_{t_\text{out}}+\ket{H}_{a^2_\text{out}}\otimes(\gamma\ket{V}+\delta\ket{H})_{t_\text{out}}).
\end{equation}
Therefore, when ancilla photon $a^2_\text{out}$ collapses onto $\ket{H}$, the target photon in $t_\text{out}$ is projected onto
\begin{equation}
\ket{t_\text{out}}=\gamma\ket{V}+\delta\ket{H},
\label{case1H}
\end{equation}
and when $a^2_\text{out}$ collapses onto $\ket{V}$, the target photon is projected onto
\begin{equation}
\sigma_x\ket{t_\text{out}}=\gamma\ket{H}+\delta\ket{V}.
\label{case2H}
\end{equation}
On the other hand, if the photon in mode $a^2_\text{in}$ is in polarization state $\ket{V}$, detecting exactly one $\ket{H}$ photon at $D_3$ results in target state
\begin{equation}
\ket{t_\text{out}}=\gamma\ket{H}+\delta\ket{V},
\label{case1V}
\end{equation}
and instead detecting exactly one $\ket{V}$ photon at $D_4$ yields
\begin{equation}
\sigma_x\ket{t_\text{out}}=\gamma\ket{V}+\delta\ket{H}.
\label{case2V}
\end{equation}
As in the case of $\text{PBS}_1$, depending on the measurement outcome for $a^2_\text{out}$, a feed-forward Pauli operation may need to be applied as indicated in (\ref{case2H}) and (\ref{case2V}). Evidently, $\text{PBS}_2$ performs a CNOT gate on modes $a^2_\text{in}$ and $t_\text{in}$, conditioned on projective measurement of the photon in $a^2_\text{out}$ (with the logic definitions of H and V swapped); the output target state of this destructive CNOT is encoded on the photon in $t_\text{out}$.
\par
Finally, assuming feed-forward operations are applied as needed, we consider the destructive CNOT gate $\text{PBS}_2$ with input
\begin{align}
&\ket{\psi_{\text{PBS}_1}}\otimes\ket{t_\text{in}}=(\alpha\ket{VH}+\beta \ket{HV})_{a^2_\text{in},c_\text{out}}\otimes(\gamma\ket{H}+\delta\ket{V})_{t_\text{in}}\nonumber\\
&=\alpha\ket{H}_{c_\text{out}}\otimes\ket{V}_{a^2_\text{in}}\otimes(\gamma\ket{H}+\delta\ket{V})_{t_\text{in}}+\beta\ket{V}_{c_\text{out}}\otimes\ket{H}_{a^2_\text{in}}\otimes(\gamma\ket{H}+\delta\ket{V})_{t_\text{in}}.
\end{align}
Based on (\ref{case1H})-(\ref{case2V}), the gate output is evidently
\begin{equation}
\ket{c_{\text{out}},t_{\text{out}}}=\alpha\ket{H}\otimes(\gamma\ket{H}+\delta\ket{V})+\beta\ket{V}\otimes(\gamma\ket{V}+\delta\ket{H}),
\end{equation}
as desired.

\subsection*{Data analysis}
\label{sec:dataAnalysis}
To characterize the CNOT gate's operation, we analyzed the control-target output photons in different polarization bases and counted four-fold coincidence detection events, to determine the output for each input state we considered. Such events can be produced in three different ways: (i) the control-target source can input two pairs, (ii) the ancilla source can input two pairs, or (iii) each source can input one pair, as desired. Since the probability of either source producing one pair during a single laser pulse is small, the probabilities of these three types of emissions are approximately equal, and hence the higher-order-emission noise can even be larger than the signal of interest. Therefore, we subtract from all measured four-fold coincidence rates the rates with the control-target source blocked, and the rates with the ancilla source blocked. It is interesting to note that the gate architecture intrinsically suppresses part of this noise: if both sources produce pairs with 100\% fidelity, albeit probabilistically, then two-photon interference makes it impossible for double-pair emission from the ancilla source to produce a four-fold detection. As discussed in \textbf{\nameref{sec:sourceQuality}}, we cannot take full advantage of this simplification in practice because of imperfections in our source.
The truth table and reconstructed density matrices with and without noise subtraction are given in the Supplementary Material.
\par
For the computational-basis case, we measured the control and target outputs in the computational basis for each of the four computational-basis input states. We simulated feed-forward in post-processing: for each four-fold coincidence, we recorded the control-target output state that would have resulted after one of the four Pauli rotations $\mathbb{I}_2\otimes\mathbb{I}_2,\mathbb{I}_2\otimes\sigma_x,\sigma_z\otimes\mathbb{I}_2,\sigma_z\otimes\sigma_x$, depending on which of the four possible ancilla detection outcome combinations occurred. We inferred the truth table in Fig. \ref{fig:bellstates} by assuming the probability of measuring each output state was proportional to the corresponding count rate observed. For computational-basis inputs, it is impossible for a double-pair emission from the control-target source to produce a four-fold coincidence, so it was only necessary to measure and subtract the rate of double-pair emission from the ancilla source.
\par
To demonstrate the gate's superposition-basis operation and generation of entanglement, measurements in a single polarization basis do not suffice. Therefore, for the case of control-target input state $\ket{D,H}$, we counted four-fold events for the three mutually unbiased analyzer settings $H,D,L$ for control and target, a total of nine two-photon measurement settings. We again assumed that the probability of each measurement outcome was proportional to the corresponding count rate. For this case, we did not simulate feed-forward: for the four possible ancilla outcomes, we reconstructed the corresponding four density matrices in Fig. \ref{fig:bellstates} using maximum-likelihood quantum state tomography \cite{tomography}. In the superposition-basis case, a double-pair emission from either source is as likely to produce a four-fold coincidence as one pair from each, so it is necessary to measure and subtract both sources' higher-order emission rates. 
\par
To determine the errors on the inferred output states, we assumed Poissonian statistics for all measured count rates, including the higher-order-emission noise rates, and performed Monte Carlo analysis with 1000 randomly generated samples.
\par

\subsection*{SPDC source quality}
\label{sec:sourceQuality}
Even for perfect gate operation, the output state fidelity is naturally limited by the quality of the input state $\ket{c_\text{in},t_\text{in}}\otimes\ket{a^1_\text{in},a^2_\text{in}}$. Our SPDC sources emit high-fidelity polarization states: quantum state tomography of the entangled input ancilla state yielded $(94.5\pm 1) $\% fidelity with the Bell state $\ket{\Psi^{-}}$. It is important to note that this fidelity is achieved at $\SI{612(20)}{mW}$ pump power. The error is dominated by coupling drifts over the duration of a measurement. 

The largest decrease in gate performance is caused by the inherent frequency entanglement of the signal-idler pairs produced by each SPDC source. We use narrowband filters to reduce these correlations, but must ultimately strike a balance between spectral unentanglement and reasonable count rates. To quantify the spectral distinguishability of our photons, we interfered each of the signal and idler photons from the control-target source with each of the signal and idler photons from the ancilla source at a beam splitter (first passing all photons through polarizers at $H$), and recorded the Hong-Ou-Mandel visibilities for all four combinations. After subtracting higher-order noise, we measured the visibilities to be
\begin{equation}
V=\frac{C_{max}-C_{min}}{C_{max}}=0.88 \pm 0.05.
\end{equation} 
and $V=0.77\pm 0.05$ without subtracting higher order noise.
As a result of this finite visibility, the gate fidelity we present is limited. Additionally, double-pair emission from the ancilla source can lead to four-fold coincidences because the two-photon interference that would normally suppress such events is imperfect (see \textbf{\nameref{sec:dataAnalysis}}).

\newpage
\subsection*{Data availability}
The authors declare that the main data supporting the finding of this study
are available within the article and its Supplementary Information.
Additional data can be provided upon request.

\subsection*{Acknowledgements}
We want to thank Sarah E. Stoeckl for contributing to the project by designing the detection stage rack. 
P.W. acknowledges support from the European Commission through EQUAM (No. 323714), PICQUE (No. 608062) and QUCHIP (No. 641039), and from the Austrian Science Fund (FWF) through START (Y585-N20), CoQuS (W1210-4) and NaMuG (P30067-N36), and the U.S. Air Force Office of Scientific Research (FA2386-17-1-4011).
A.S. acknowledges support from the German Research foundation (SZ 276/12-1 and BL 574/13-1).

\subsection*{Competing interests}
The authors declare no competing financial interests.

\subsection*{Author contributions}
J.Z., M.T. designed the experiment. J.Z, A.N.S., M.T. built the setup and carried out data collection, J.Z., A.N.S. performed data analysis, R.H., M.G. designed and fabricated the waveguide, A.M. proposed the use of TEC fibers, P.W. and A.S. supervised the project. All authors contributed to writing the paper.


\printbibliography

\end{document}